\documentclass[10pt,onecolumn,twoside,draftcls]{IEEEtran}

\usepackage{tikz}
\usepackage{blindtext}
\usepackage{graphicx}
\usepackage{amsmath}
\usepackage{mathrsfs}
\usepackage[sans]{dsfont}
\usepackage{amsbsy}
\usepackage{amssymb}
\usepackage{pdftexcmds}
\usepackage{enumerate}
\usepackage[capitalise]{cleveref}
\usepackage{cite}
\usepackage{accents}

\newcommand{\ubar}[1]{\underaccent{\bar}{#1}}

\DeclareMathOperator{\Equaldef}{\overset{def}{=}}
\renewcommand{\Pr}{\mathbf{P}}

\newtheorem{problem}{Problem}
\newtheorem{proposition}{Proposition}
\newtheorem{remark}{Remark}
\newtheorem{theorem}{Theorem}
\newtheorem{definition}{Definition}

\newtheorem{lemma}{Lemma}

\newcommand{\diag}{\mathop{\mathrm{diag}}}
\newcommand{\mean}{\mathop{\mathrm{mean}}}

\title{Observation-driven scheduling for remote estimation of two Gaussian sources}

\author{Marcos M. Vasconcelos and Urbashi Mitra\thanks{M. M. Vasconcelos and U. Mitra are with the Department of Electrical Engineering, University of Southern California, Los Angeles, CA 90089 USA. E-mails: \texttt{\{mvasconc,ubli\}@usc.edu}. This work was supported in part by the following agencies: ONR under grant N00014-15-1-2550, NSF under grants CNS-1213128, CCF-1718560, CCF-1410009, CPS-1446901 and AFOSR under grant FA9550-12-1-0215.  }}

\begin{document}

\maketitle

\begin{abstract}
Joint estimation and scheduling for sensor networks is considered in a system formed by two sensors, a scheduler and a remote estimator. Each sensor observes a Gaussian source, which may be correlated. The scheduler observes the output of both sensors and chooses which of the two is revealed to the remote estimator. The goal is to jointly design scheduling and estimation policies that minimize a mean-squared estimation error criterion. The person-by-person optimality of a policy pair called ``max-scheduling/mean-estimation'' is established, where the measurement with the largest absolute value is revealed to the estimator, which uses a corresponding conditional mean operator. This result is obtained for independent sources, and in the case of correlated sources and symmetric variances. We also consider the joint design of scheduling and linear estimation policies for two correlated Gaussian sources with an arbitrary correlation structure. In this case, the optimization problem can be cast a difference-of-convex program, and locally optimal solutions can be efficiently found using a simple numerical procedure.  
\end{abstract}

\section{Introduction}

The multiple components of Cyber-physical systems are often interconnected by shared communication links of limited bandwidth \cite{Antsaklis:2014}. One way to model this bandwidth constraint is to assume that, at any time instant, a single packet can be reliably transmitted over the link to its destination \cite{Vasconcelos:2017}. Therefore, the system designer must come up with rules/algorithms that allocate shared communication resources among multiple transmitting nodes. This paper introduces a new class of remote estimation problems where the communication resources are allocated \textit{dynamically} based on the observations at the sensors, rather than based purely on the statistical description of the sources.

\begin{figure}[!t]
    \begin{center}
    \includegraphics[scale=0.45]{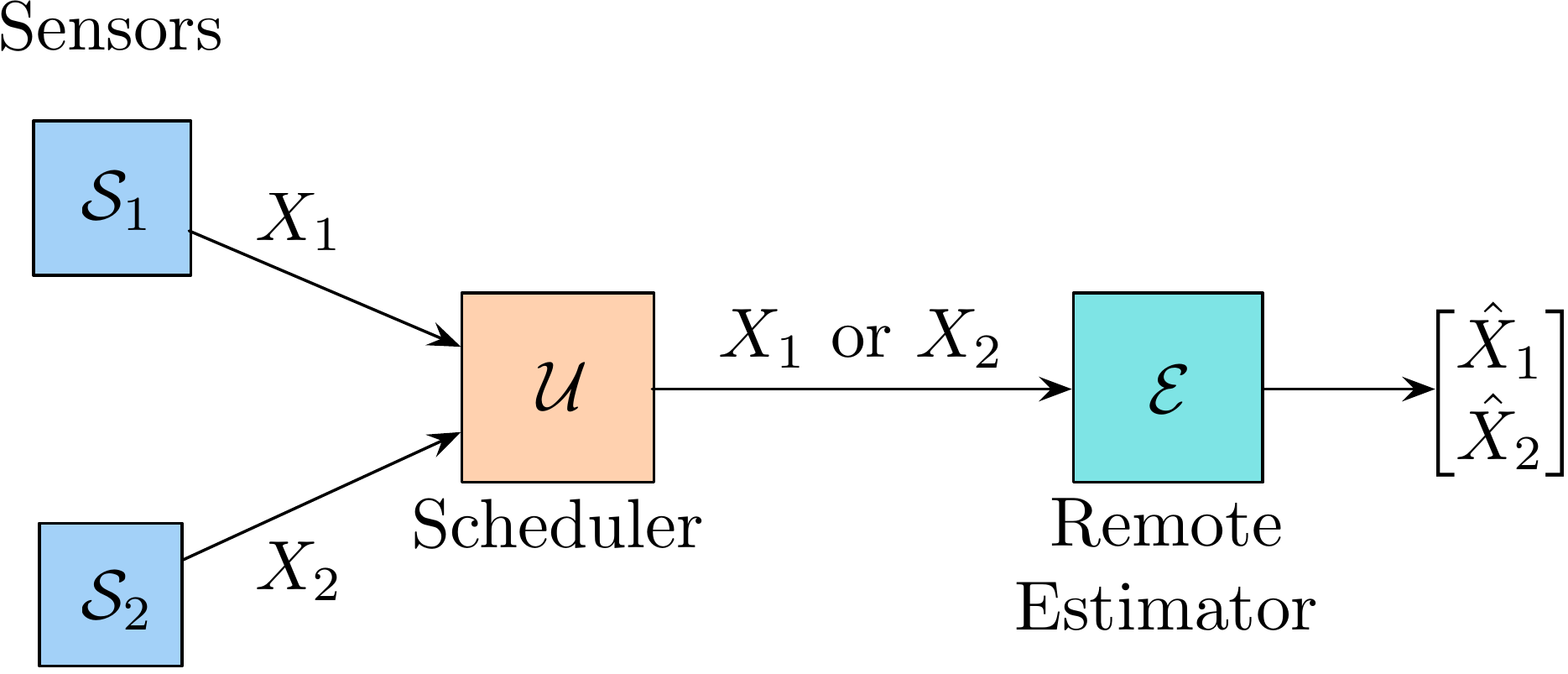}
\end{center}
\caption{Block diagram of the observation-driven sensor scheduling problem.}
\label{fig:system}
\end{figure}

The basic framework considered is shown in \cref{fig:system}. Two sensors, possibly making correlated observations, report their measurements to a scheduler. The role of the scheduler is to select one of the observations and transmit it to a remote estimator. Finally, the remote estimator forms estimates of both measurements. Our goal is to jointly design scheduling and estimation policies that minimize a mean-squared estimation error. Alternatively, this problem can be understood as one of dimensionality reduction \cite{Roweis:2000}, where an encoder-decoder pair is designed to minimize the expected distortion between the original and reconstructed vectors, with the constraint that a scalar (versus a vector) is transmitted or stored. The solution to this canonical problem formulation can be used to drive the design of scheduling algorithms for more complex networked systems, where decisions on what is transmitted or not are made in real-time.


The problem of selecting a subset among a larger set of sensors with the purpose of transmission over a bandwidth constrained network and subsequent estimation is generally referred to as \textit{sensor scheduling}, with applications spanning many areas in engineering such as networked control, sensor networks, target tracking and remote estimation \cite{Hespanha:2007,Akyildiz:2002,Ny:2011}. This class of problems has a long and rich history initiated with \cite{Athans:1972}. In general, sensor scheduling is a hard combinatorial optimization problem \cite{Moon:2017}. However, the computational complexity may be circumvented by suboptimal pruning of decision trees \cite{Hovareshti:2007,Vitus:2012}. Another approach to address the complexity issue is by use convex relaxations \cite{Joshi:2009,Li:2015}. In certain cases, it is possible to show that the solution to these relaxed problems yield optimal scheduling schemes which are periodic and therefore admit simple implementations \cite{Mo:2011,Shi:2012}. In a related line of work, a framework for sensor selection where a cost function augmented with a sparsity promoting term is introduced with the goal of trading off complexity vs. performance  \cite{Dhingra:2014}. 


Our approach to the scheduling problem is aligned with the work of \cite{Xia:2017,Vasconcelos:2017c} where the decision is made based on the realizations of the measurements themselves. The idea is to design and exploit \textit{event-triggers} \cite{Wu:2013} for the transmission of one of the variables over the other, which allows for implicit communication via signaling \cite{Ho:1978}. In a way, the problem we address here is an \textit{observation selection} problem, such as in \cite{Krause:2007} and the techniques we use in the design of decision making policies are reminiscent of quantization theory \cite{Gray:2006}, where the observation space is partitioned in regions where certain decisions are made. In the context of our problem, the observations ``trigger'' which one of the measurements is transmitted over the communication link. Interestingly, in our scheduling/estimation schemes, the transmitted variable is used as side information for the estimation of the non-transmitted variable. 


The problem addressed in this paper is directly related to the infamous ``Witsenhausen's Counterexample''  and the ``Gaussian Test Channel'' \cite{Witsenhausen:1968,Basar:2008}. The connection between these classical problems and ours is that the joint design of scheduling and estimation policies is entangled by signaling. In other words, the action of the scheduler directly affects what the estimator observes, which turns this problem one of team-decision with a non-classical information structure \cite{Yuksel:2013}.

Finally, the problem studied in this paper is closely related to the problem of estimating random variables observed by individual sensors, which independently decide to transmit over a collision channel \cite{Vasconcelos:2017}. In that case, unless a sensor uses a policy to remain always silent, collisions may occur. Here, the inclusion of a scheduler has the goal of completely avoiding collisions. In a sense, the problem considered here is a ``centralized'' version of the problem in \cite{Vasconcelos:2017}, which in principle can be used to lower bound the performance of the decentralized system.


\subsection{Contributions and Organization}

The main contributions of this paper are:

\begin{itemize}

	\item We establish the \textit{person-by-person optimality} of the max-scheduling/mean-estimation policy pair for sensors making independent Gaussian observations. One remarkable feature of this result is that the structure of the scheduling policy is completely independent of the variances of the observations. The mean estimation policy, in this case, is a piecewise linear function of received packet at the remote estimator.

	\item We establish the \textit{person-by-person optimality} of the same pair of strategies in the case when the observations are \textit{correlated} but have equal variances. In this case, the mean-estimation policy is a non-linear function of the information received by the sensor over the channel. The proof of this result depends on the symmetry and monotonicity properties related to a soft-thresholding nonlinear estimator induced by the max-scheduling policy. 

	\item Our third contribution is to provide a numerical procedure that efficiently solves a non-convex optimization problem when the estimators are constrained to the class of piece-wise linear functions. For two sensors, the solutions found by this algorithm can be verified to be globally optimal.

	\item Finally, we extend the person-by-person optimality result to account for any number of sensors observing independent zero mean Gaussian sources.

\end{itemize}

Preliminary versions of the Theorems 1 and 2 presented here have appeared previously in \cite{Vasconcelos:2017b}, where certain key technical aspects of the proofs were either conjectured or omitted. The proofs of the results reported here are detailed and precise. Additionally, we provide several new results which have not appeared elsewhere in Theorems 3, 4, 5 and 6. Another important key contribution is the derivation of an efficient numerical algorithm for the design of piecewise linear minimum mean squared error estimators.


The article is organized in nine sections including the Introduction. In Section II, we state the precise problem formulation and define the two notions of optimality which are used throughout the paper. Then, we state the two main theoretical results of the paper in Section III. The proof of \Cref{thm:pbp}, which concerns the case of independent Gaussian observations is presented in Section IV and the proof of \Cref{thm:pbp_corr} for the correlated case with symmetric variances is presented in Section V. In Section VI, we provide another person-by-person optimality result that addresses the case of general covariance matrix structure by using a linear \textit{decorrelating transform}. In the last part of the paper, the optimization problem is constrained to the class of piecewise linear estimators. In Section VII, we obtain locally optimal solutions for the case with a general covariance matrix using a numerical procedure based on the Convex-Concave Procedure. Finally, we extend the result on the independent case to an arbitrary number of sensors observing Gaussian random variables in Section VIII. We conclude in Section IX with our final remarks and suggestions for future work. 

\subsection{Notation}

We adopt the following notation: random variables and random vectors are represented using upper case letters, such as $X$. Realizations of random variables and random vectors are represented by the corresponding lower case letter, such as $x$.  The probability density function of a continuous random variable $X$, provided that it is well defined, is denoted by $f_X$. Functions and functionals are denoted using calligraphic letters such as $\mathcal{F}$. We use $\mathcal{N}(m,\sigma^2)$ to represent the Gaussian probability distribution of mean $m$ and variance $\sigma^2$, respectively. The real line is denoted by $\mathbb{R}$. Sets are represented in blackboard bold font, such as $\mathbb{A}$. The probability of an event $\mathfrak{E}$ is denoted by $\Pr(\mathfrak{E})$; the expectation of a random variable $Z$ is denoted by $\mathbf{E}[Z]$. The indicator function of a statement $\mathfrak{S}$ is defined as follows: 
\begin{equation}
\mathbf{1}\big(\mathfrak{S}\big) \Equaldef \begin{cases}
1 & \text{if} \ \ \mathfrak{S}\ \  \text{is true}\\
0 & \text{otherwise}.
\end{cases}
\end{equation}


\section{Problem formulation}

Consider the system in \cref{fig:system} comprised of two sensors labeled $\mathcal{S}_1$ and $\mathcal{S}_2$. Each sensor observes a Gaussian random variable with known mean and variance. Without loss of generality, we assume that sensor $\mathcal{S}_i$ observes $X_i$, where
\begin{equation}
X_i \sim \mathcal{N}(0,\sigma_i^2), \ \ i \in\{1,2\}.
\end{equation}
The correlation coefficient between $X_1$ and $X_2$ is defined as:
\begin{equation}
\rho\Equaldef \frac{\mathbf{E}[X_1X_2]}{\sigma_1\cdot\sigma_2}.
\end{equation}

The observations $X_1$ and $X_2$ must be communicated to a remote estimator over a communication link, where a single packet is transmitted to the remote estimator at a time.




The scheduler's decision variable, denoted by $U$, is computed according to a \textit{scheduling policy}, which is a measurable function $\mathcal{U}: \mathbb{R}^2 \rightarrow \{1,2\}$ such that 
\begin{equation}
U=\mathcal{U}(X_1,X_2).
\end{equation}
The set of all admissible scheduling policies is denoted by $\mathbb{U}$.

The scheduler's decision $U$ determines what the remote estimator observes as follows:
\begin{equation}
Y=(U,X_U).
\end{equation}
The vector $Y$ belongs to the set $\mathbb{Y}\Equaldef \{1,2\} \times \mathbb{R}$.
\begin{remark}
Notice that the scheduler effectively sends a packet containing the index $U$ in addition to the real number $X_U$. The reason behind this assumption is to let the estimator know the origin of the packet before forming its estimates. The presence of an identification number on a packet is a standard assumption in data networks \cite{Bertsekas:1992}. 
\end{remark}

Upon observing $Y$, the remote estimator forms estimates of the observations at both sensors $X_1$ and $X_2$, denoted by $\hat{X}_1$ and $\hat{X}_2$, respectively. This is done according to an estimation policy $\mathcal{E}: \mathbb{Y}\rightarrow \mathbb{R}^2$ as follows:
\begin{equation}
( \hat{X}_1 , \hat{X}_2) = \mathcal{E}(Y).
\end{equation}
The set of all admissible estimation policies is denoted by $\mathbb{E}$.

Our goal is to solve the following optimization problem.

\begin{problem}\label{prob:main}
Given the variances $\sigma_1^2, \sigma_2^2 >0$ and the correlation coefficient $\rho \in [0,1)$, find a scheduling and estimation policy pair $(\mathcal{U},\mathcal{E})\in \mathbb{U} \times \mathbb{E}$ that jointly minimizes the following cost:
\begin{equation}{\label{eq:cost}}
\mathcal{J}(\mathcal{U},\mathcal{E}) \Equaldef \mathbf{E} \left[ \big( X_1-\hat{X}_1\big)^2+\big( X_2-\hat{X}_2\big)^2\right].
\end{equation}
\end{problem}

\subsection{Notions of optimality}

\subsubsection{Global optimality} A pair of scheduling and estimation strategies $(\mathcal{U}^{\star},\mathcal{E}^{\star}) \in \mathbb{U}\times\mathbb{E}$ is \underline{globally optimal} if
\begin{equation}
\mathcal{J}(\mathcal{U}^{\star},\mathcal{E}^{\star}) \leq 
\mathcal{J}(\mathcal{U},\mathcal{E}), \ \ (\mathcal{U},\mathcal{E}) \in \mathbb{U}\times\mathbb{E}.
\end{equation}

\subsubsection{Person-by-person optimality}
A pair of scheduling and estimation strategies $(\mathcal{U}^{\star},\mathcal{E}^{\star}) \in \mathbb{U}\times\mathbb{E}$ is \\ \underline{person-by-person optimal} if
\begin{IEEEeqnarray}{rCl}
\mathcal{J}(\mathcal{U}^{\star},\mathcal{E}^{\star}) &\leq& 
\mathcal{J}(\mathcal{U},\mathcal{E}^{\star}), \ \ \mathcal{U} \in \mathbb{U}\\
\mathcal{J}(\mathcal{U}^{\star},\mathcal{E}^{\star}) &\leq& 
\mathcal{J}(\mathcal{U}^{\star},\mathcal{E}), \ \ \mathcal{E} \in \mathbb{E}.
\end{IEEEeqnarray}

\section{Main results}

The main contribution of this work is to establish the person-by-person optimality of several pairs of scheduling and estimation policies for \cref{prob:main} for the different structures of correlation between the observations $X_1$ and $X_2$. Before formally stating the results, we first define the max-scheduling, mean-estimation, and soft-thresholding estimation policies.

\begin{definition}[max-scheduling policy]
Let $x \in \mathbb{R}^2$. The max-scheduling policy is defined as:
\begin{equation}
\mathcal{U}^{\max}(x) \Equaldef \begin{cases}
1 & \text{if} \ \ |x_1| \geq |x_2|\\
2 & \text{otherwise}.
\end{cases}
\end{equation}
\end{definition}

\begin{definition}[mean-estimation policy]
Let $\xi \in \mathbb{R}$. The mean-estimation policy is defined as:
\begin{equation}
\mathcal{E}^{\mathrm{mean}}(i,\xi) = \begin{cases} \big[ \ \xi \ \  0 \ \big]^\mathsf{T} & \text{if} \ \ i=1 \\
\big[ \ 0 \ \  \xi \ \big]^\mathsf{T} & \text{if} \ \ i=2. 
\end{cases}
\end{equation}
\end{definition}
The reason why the policy above is called \textit{mean-estimation} is that the estimator outputs the \textit{mean} of the unobserved random variable as an estimate. In other words, the side information provided by observing $X_i=\xi$ is irrelevant for estimating $X_j$, $i\neq j$.  In this case, since the random variables $X_1$ and $X_2$ are assumed to be zero-mean, the mean-estimation policy takes the form above. 

\begin{definition}[soft-thresholding estimation policy]
The soft-thresholding estimation policy is defined as:
\begin{equation}
\mathcal{E}^{\mathrm{soft}}(i,\xi) = \begin{cases} \big[ \ \xi  \ \ \eta(\xi) \ \big]^\mathsf{T} & \text{if} \ \ i=1 \\
\big[ \ \eta(\xi)  \ \ \xi \ \big]^\mathsf{T} & \text{if} \ \ i=2. 
\end{cases}
\end{equation}
where $\eta(\xi)$ is a nonlinear soft-thresholding function with parameters $\sigma^2>0$ and $\rho\in [0,1)$ defined as:
\begin{equation}\label{eq:nonlinear}
\eta(\xi) \Equaldef  \frac{\int_{-|\xi|}^{|\xi|} \tau \exp\left(-\frac{(\tau-\rho \xi)^2}{2\sigma^2(1-\rho^2)}\right) d\tau }{\int_{-|\xi|}^{|\xi|} \exp\left(-\frac{(\tau-\rho \xi)^2}{2\sigma^2(1-\rho^2)}\right) d\tau }.
\end{equation}
\end{definition}

Our main results are stated in the Theorems bellow.

\begin{theorem}\label{thm:pbp}
If $\rho=0$, the policy pair $(\mathcal{U}^{\max},\mathcal{E}^{\mean})$ is a person-by-person optimal solution for the cost $\mathcal{J}(\mathcal{U},\mathcal{E})$ in \cref{eq:cost}.
\end{theorem}

\begin{theorem}\label{thm:pbp_corr}
If $\sigma^2_1=\sigma^2_2$, the policy pair $(\mathcal{U}^{\max},\mathcal{E}^{\mathrm{soft}})$ is a person-by-person optimal solution for the cost $\mathcal{J}(\mathcal{U},\mathcal{E})$ in \cref{eq:cost}.
\end{theorem}

\begin{remark}
\Cref{thm:pbp,thm:pbp_corr} present candidates for globally optimal scheduling and estimation policy pairs for \Cref{prob:main}. We conjecture that these pairs are globally optimal. However, at this point there are no analytical tools to make stronger statements. An alternate way to interpret this result is from the perspective from game theory, as \Cref{thm:pbp,thm:pbp_corr} say that the pairs $(\mathcal{U}^{\max},\mathcal{E}^{\mean})$ and $(\mathcal{U}^{\max},\mathcal{E}^{\mathrm{soft}})$ constitute Nash-equilibrium solutions \cite{Basar:1999}. 
\end{remark}

\section{Independent observations}

We start with the simpler case where the sensors make independent measurements. Let $X_1$ and $X_2$ be uncorrelated scalar Gaussian random variables, i.e., the correlation coefficient $\rho=0$.
We will now state two necessary optimality conditions reminiscent of quantization theory \cite{Gersho:1992}. The first property pertains to the optimality of an optimal estimation policy for an arbitrarily fixed scheduling policy $\mathcal{U} \in \mathbb{U}$.    

\begin{lemma}[Optimal estimator]\label{lem:estimator}
For a fixed scheduling policy $\mathcal{U} \in \mathbb{U}$, the estimation policy that minimizes the mean squared error cost in \cref{eq:cost} is the following:
\begin{equation}
\mathcal{E}^{\star}_{\mathcal{U}} (y) = \mathbf{E}\left[X \mid Y=y\right].
\end{equation}
\end{lemma}

\begin{IEEEproof}[Proof citation]
This is the classical nonlinear filtering result. Its proof is found in many texts, such as \cite[pg. 143]{Poor:1998}.
\end{IEEEproof}
{}
\begin{remark}
There are two noteworthy facts about \Cref{lem:estimator}: ($i$) The optimal estimation policy is always a function of the scheduling policy. This coupling leads to the lack of convexity of \Cref{prob:main}; ($ii$) The scheduling policy creates a coupling between the random variables $X_1$ and $X_2$ even when they are independent, which means that no matter what is received by the remote estimator should be used as side information for forming the optimal estimates $\hat{X}_1$ and $\hat{X}_2$.
\end{remark}

\begin{lemma}[Identity structure]\label{lem:estimator_identity}
The search for optimal estimation policies can be constrained to the set of policies $\mathcal{E}$ that satisfy the following identity property: 
\begin{equation}
\mathcal{E}(1,\xi) = \begin{bmatrix} \xi \\ \eta_2(\xi) \end{bmatrix} \ \ \text{and} \ \ \mathcal{E}(2,\xi) = \begin{bmatrix}  \eta_1(\xi) \\ \xi \end{bmatrix},
\end{equation}
where $\eta_i:\mathbb{R} \rightarrow \mathbb{R}$, $i\in \{1,2\} $. 
\end{lemma}

\begin{IEEEproof}
Let $i,j \in\{1,2\}$ such that $i\neq j$, then for any fixed scheduling policy $\mathcal{U} \in \mathbb{U}$ the event $\big\{Y=(i,\xi)\big\}$ is equivalent to the event $\big\{U=i,X_i=\xi\big\}$. Therefore,
\begin{IEEEeqnarray}{rCl}
\mathbf{E} \big[X_i \mid Y=(i,\xi)\big] & = & \xi.
\end{IEEEeqnarray}
Similarly,
\begin{IEEEeqnarray}{rCl}
\mathbf{E}\big[ X_j \mid Y=(i,\xi)\big] & = & \int_{\mathbb{R}} x_j f_{X_j\mid U=i,X_i=\xi}(x_j)dx_j.
\end{IEEEeqnarray}
\end{IEEEproof}

For the remainder of this article, every admissible estimator $\mathcal{E}\in \mathbb{E}$ satisfies the identity property in \Cref{lem:estimator_identity} and therefore is completely specified by so-called \textit{representation functions} denoted by $\eta_1$ and $\eta_2$. This fact will be used to establish a necessary optimality condition for the optimal scheduling policy, for a given estimation policy $\mathcal{E}\in \mathbb{E}$.

\begin{lemma}[Generalized nearest neighbor condition]\label{lem:scheduler}
For a fixed estimation policy $\mathcal{E} \in \mathbb{E}$ parameterized by representation functions $\eta_1$ and $\eta_2$, the following scheduling policy minimizes the cost in \cref{eq:cost}:
\begin{equation}
\mathcal{U}_{\mathcal{E}}^{\star} (x) = \begin{cases}
1 & \text{if} \ \ |x_1-\eta_1(x_2)| \geq |x_2-\eta_2(x_1)|  \\
2 & \text{otherwise.}
\end{cases}
\end{equation}
\end{lemma}

\begin{IEEEproof}
Using the law of total expectation, we write:
\begin{equation}
\mathcal{J}(\mathcal{U},\mathcal{E}) = \mathbf{E}\left[\|X-\hat{X}\|^2 \mid U=1\right]\Pr(U=1) 
+ \mathbf{E}\left[\|X-\hat{X}\|^2 \mid U=2\right]\Pr(U=2). 
\end{equation}
Due to the identity structure in \Cref{lem:estimator_identity}, the following holds: 
\begin{equation}
\mathcal{J}(\mathcal{U},\mathcal{E})=\int_{\mathbb{R}^2}\big(x_2-\eta_2(x_1)\big)^2\mathbf{1}\big(\mathcal{U}(x)=1\big)f_X(x)dx  + \int_{\mathbb{R}^2}\big(x_1-\eta_1(x_2)\big)^2\mathbf{1}\big(\mathcal{U}(x)=2\big)f_X(x)dx.
\end{equation}
For fixed representation functions $\eta_1$ and $\eta_2$, we can construct a scheduling policy that minimizes the expression above. Let $\mathbb{Q}_i$ be defined as:
\begin{equation}
\mathbb{Q}_i \Equaldef \big\{x \in \mathbb{R}^2 \mid \mathcal{U}(x) = i \big\}, \ \ i\in \{1,2\}.
\end{equation}
Assign to $\mathbb{Q}_1$ the points $x \in \mathbb{R}^2$ which satisfy the following inequality:
\begin{equation}
\big(x_2-\eta_2(x_1)\big)^2 \leq \big(x_1-\eta_1(x_2)\big)^2,
\end{equation}
and the remaining points are assigned to $\mathbb{Q}_2$. 
\end{IEEEproof}

\begin{remark}
Notice that \Cref{lem:scheduler} is completely independent of the joint probability density function $f_X$.
\end{remark}

We are now equipped to prove \cref{thm:pbp}.

\begin{IEEEproof}[Proof of \Cref{thm:pbp}]
Let the estimation policy be $\mathcal{E}=\mathcal{E}^{\mean}$. The associated representation functions are given by:
\begin{equation}
\eta_i (\xi) =0, \ \ \xi \in\mathbb{R}, \ \ i \in \{1,2\}. 
\end{equation}
From \Cref{lem:scheduler}, an optimal scheduling policy for $\mathcal{E}^{\mean}$ is:
\begin{equation}
\mathcal{U}_{\mathcal{E}^{\mean}}^{\star} (x) = \begin{cases}
1 & \text{if} \ \ |x_1| \geq |x_2|  \\
2 & \text{otherwise,}
\end{cases}
\end{equation}
which is equal to the max-scheduling policy $\mathcal{U}^{\max}$. 

Conversely, assume that the scheduling policy $\mathcal{U}=\mathcal{U}^{\max}$. \Cref{lem:estimator} implies that the optimal estimator is given by:
\begin{equation}
\mathcal{E}^{\star}_{\mathcal{U}^{\max}}(i,\xi) = \mathbf{E}\big[ X \mid Y=(i,\xi)\big],
\end{equation}
where $i\in\{1,2\}$ and $\xi\in \mathbb{R}$.

If $i=1$, then:
\begin{equation}
\mathbf{E}\big[X\mid Y=(1,\xi)\big] = \begin{bmatrix} \xi \\ \eta_2(\xi) \end{bmatrix},
\end{equation}
where
\begin{equation}
\eta_2(\xi) = \frac{\int_{\mathbb{R}}x_2\mathbf{1}\big(\mathcal{U}^{\max}(\xi,x_2)=1\big)f_{X_2|X_1=\xi}(x_2)dx_2}{\int_{\mathbb{R}}\mathbf{1}\big(\mathcal{U}^{\max}(\xi,x_2)=1\big)f_{X_2|X_1=\xi}(x_2)dx_2}.
\end{equation}
Since $\rho=0$, then the conditional probability density function is equal to: 
\begin{equation}
f_{X_2|X_1=\xi}(x_2) = \frac{1}{\sqrt{2\pi\sigma_2^2}}\exp\left(-\frac{x_2^2}{2\sigma^2_2}\right).
\end{equation}
Therefore, the representation function $\eta_2$ can be explicitly computed as: 
\begin{IEEEeqnarray}{rCl}
\eta_2(\xi) & = & \frac{\int_{-|\xi|}^{|\xi|}x_2\frac{1}{\sqrt{2\pi\sigma_2^2}}\exp\left(-\frac{x_2^2}{2\sigma^2_2}\right)dx_2}{\int_{-|\xi|}^{|\xi|}\frac{1}{\sqrt{2\pi\sigma_2^2}}\exp\left(-\frac{x_2^2}{2\sigma^2_2}\right)dx_2}.
\end{IEEEeqnarray}
Due to the even symmetry of the marginal Gaussian density around zero, we have:
\begin{equation}
\eta_2(\xi) = 0, \ \ \xi\in\mathbb{R}.
\end{equation}

Repeating the same steps for $i=2$, leads to:
\begin{equation}
\eta_1(\xi) = 0, \ \ \xi\in\mathbb{R}.
\end{equation}

Therefore, 
\begin{equation}
\mathcal{E}^{\star}_{\mathcal{U}^{\max}} = \mathcal{E}^{\mean}.
\end{equation}
\end{IEEEproof}

\subsection{An illustrative example}

For two independent Gaussian observations $X_1\sim\mathcal{N}(0,\sigma_1^2)$ and $X_2\sim\mathcal{N}(0,\sigma_2^2)$, the performance of the person-by-person optimal pair of policies $(\mathcal{U}^{\max},\mathcal{E}^{\mean})$ is given by the following formula:
\begin{equation}
\mathcal{J}(\mathcal{U}^{\max},\mathcal{E}^{\mean}) = \mathbf{E}\big[\min\{X_1,X_2\}\big].
\end{equation}

\begin{figure}[!t]
    \begin{center}  
      \includegraphics[width=0.5\textwidth]{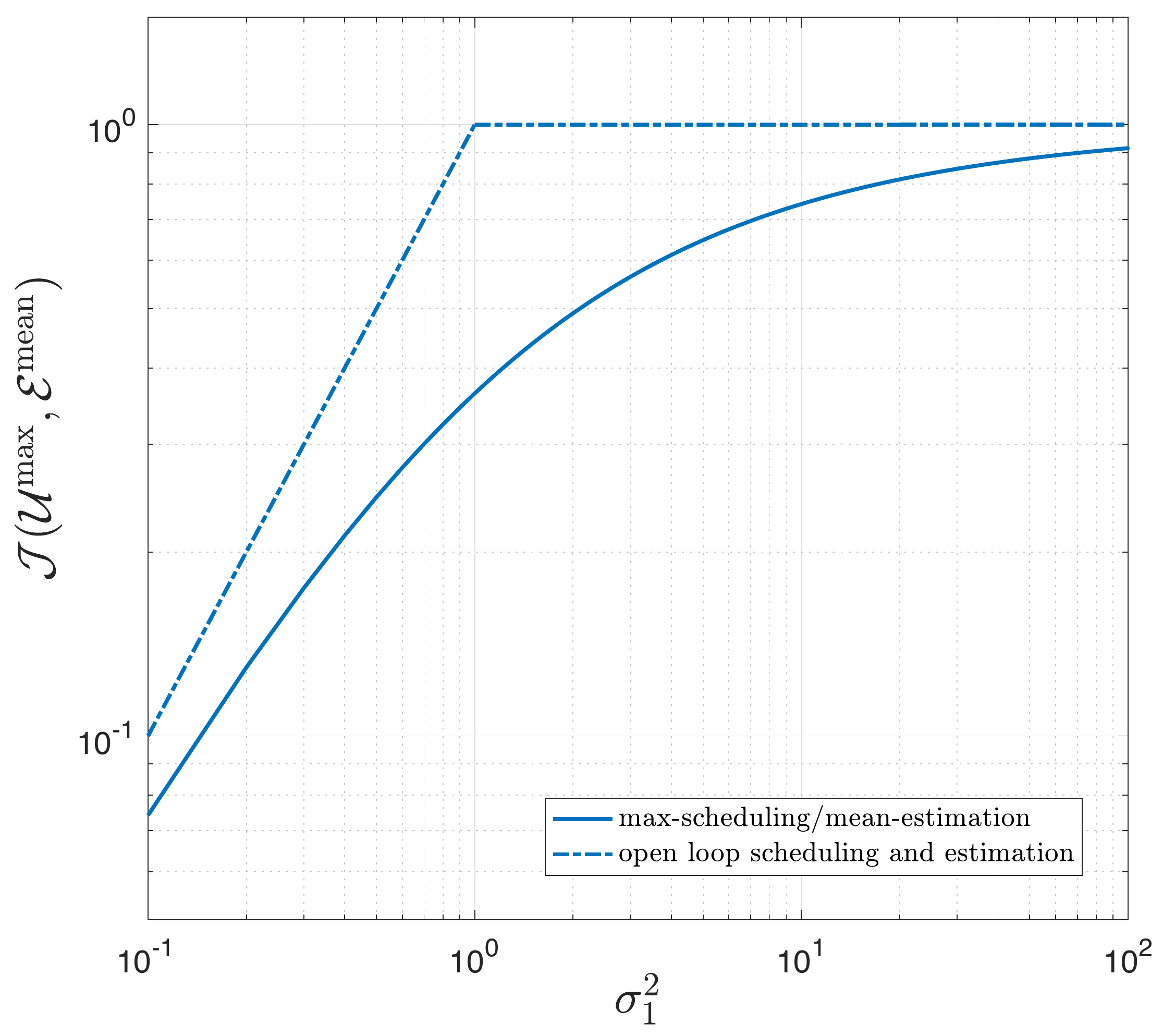}
\end{center}
\caption{Performance of the max-scheduling and mean estimation policy computed as a function of the variance $\sigma^2_1$, while keeping $\sigma^2_2=1$ fixed. The dashed plot corresponds to the performance of the open-loop scheduling policy where the sensor with largest variance is transmitted to the remote estimator.}
\label{fig:performance_indep}
\end{figure}

\Cref{fig:performance_indep} shows the performance of this pair of policies as a function of $\sigma_1^2$ while keeping $\sigma^2_2 =1$. 
In contrast with the case where the observations are not taken into account, the decision of what to transmit is based on the statistics of the source rather than the measurements. In this ``open-loop'' scheduling scheme, the source with the largest variance is always is transmitted, i.e.,
\begin{equation}
\mathcal{U}^{\mathrm{open}}(x) \Equaldef \arg \max_{i \in \{1,2\}} \sigma_i^2.
\end{equation}
Interestingly, the optimal estimator for the policy above is:
\begin{equation}
\mathcal{E}^{\star}_{\mathcal{U}^{\mathrm{open}}} = \mathcal{E}^{\mean}.
\end{equation}
Therefore, the performance of the open-loop scheme is thus given by the following expression:
\begin{equation}
\mathcal{J}(\mathcal{U}^{\mathrm{open}},\mathcal{E}^{\mathrm{mean}}) = \min\{\sigma^2_1,\sigma^2_2\}.
\end{equation}
\Cref{fig:performance_indep} shows the performance of the two schemes and the improvement achieved by the max-scheduling policy, which schedules the transmissions among sensors dynamically. The gap between the two curves is the ``value-of-information'', i.e., how much we can gain from the additional information contained in the realizations for the scheduling problem.

\section{The symmetric correlated case}

The two essential properties that enabled us to come up with a simple proof for the person-by-person optimality result in Theorem 1 were: ($i$) The fact that the two random variables $X_1$ and $X_2$ are independent; ($ii$) The fact that the (conditional) Gaussian pdfs are symmetric about the mean (which is zero in this case). When considering correlated Gaussian observations, these two properties no longer hold.

We proceed with exploring the case when the variances are equal, but the observations are correlated, i.e., the covariance matrix is:
\begin{equation}
\mathbf{\Sigma} = \sigma^2 
\begin{bmatrix}
 1 & \rho\\
 \rho & 1
\end{bmatrix}.
\end{equation}
In this case, the conditional density of $X_i \mid X_j = \xi$ is:
\begin{equation}
\mathcal{N}\left(\rho \xi,\sigma^2(1-\rho^2)\right).
\end{equation}

Let us define the optimal nonlinear representation functions induced by the max-scheduling policy when the observations are symmetrically correlated. Let $i,j \in \{1,2\}$ such that $i\neq j$. Then, under the max-scheduling policy we have:
\begin{equation}
\mathbf{E}\big[X_i\mid Y=(j,\xi)\big]  =  \frac{\int_{-|\xi|}^{|\xi|} x_i f_{X_i\mid X_j=\xi}(x_i)dx_i }{\int_{-|\xi|}^{|\xi|} f_{X_i\mid X_j=\xi}(x_i)dx_i }.
\end{equation}


Notice that, due to the symmetric variances, the two nonlinear estimates corresponding to $i=1,2$ given by the expression above are equal. This leads to the nonlinear soft-thresholding representation function:

\begin{equation}
\eta(\xi) \Equaldef  \frac{\int_{-|\xi|}^{|\xi|} \tau \exp\left(-\frac{(\tau-\rho \xi)^2}{2\sigma^2(1-\rho^2)}\right) d\tau }{\int_{-|\xi|}^{|\xi|} \exp\left(-\frac{(\tau-\rho \xi)^2}{2\sigma^2(1-\rho^2)}\right) d\tau }.
\end{equation}

\begin{figure}[!t]
    \begin{center}  
      \includegraphics[width=0.5\textwidth]{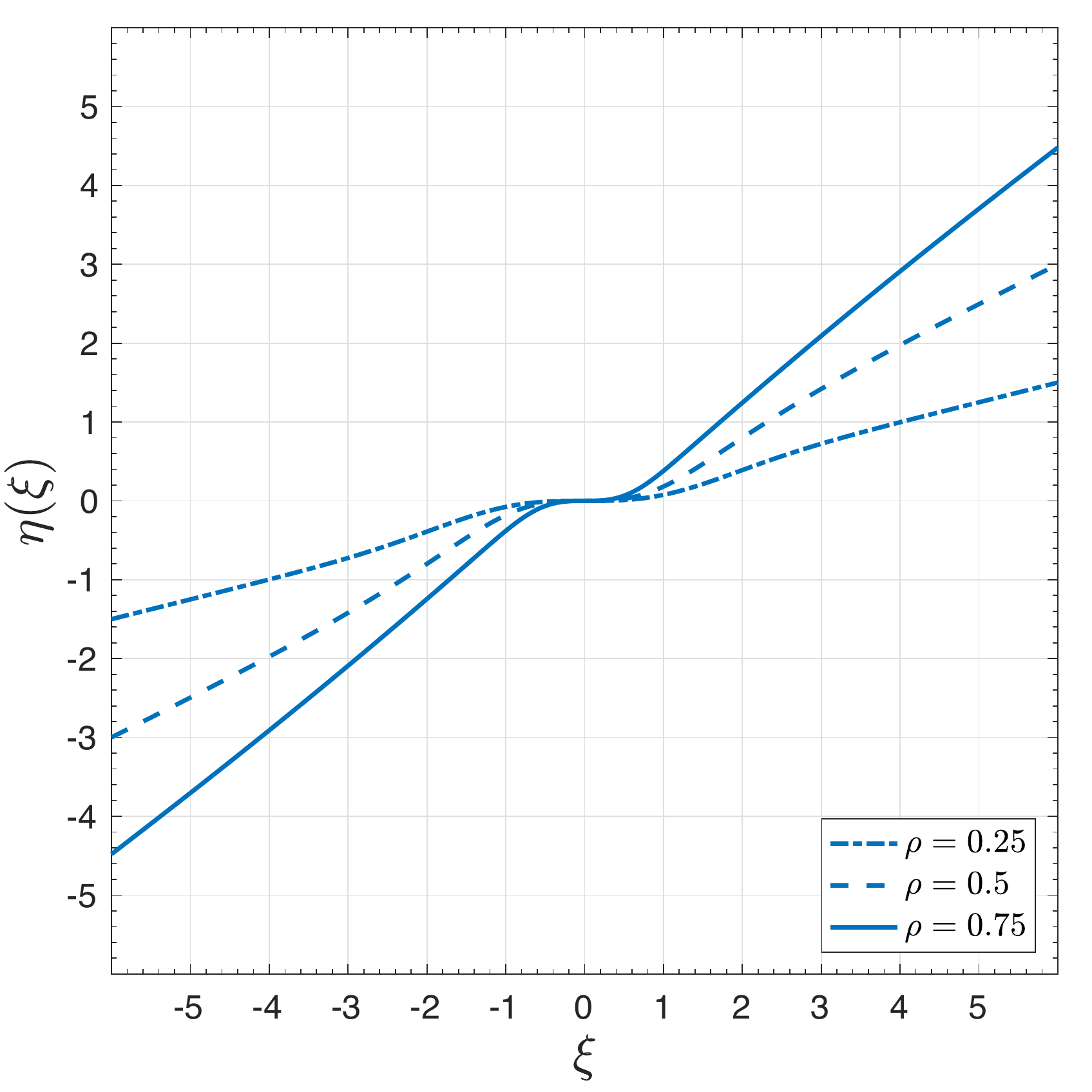}
\end{center}
\caption{Nonlinear soft-thresholding estimator induced by the max-scheduling policy for symmetric Gaussian sources. In this figure the variance is normalized to one.}
\label{fig:nonlinear}
\end{figure}

The representation function $\eta(\xi)$ is shown in \cref{fig:nonlinear}. It is straightforward to show that $\eta$ has odd symmetry. We state this fact without proof as a lemma. 

\begin{lemma}[Odd symmetry of the nonlinear soft-thresholding representation function]\label{lem:odd}
The function $\eta$ defined in \cref{eq:nonlinear} satisfies:
\begin{equation}
\eta(-\xi) = - \eta(\xi), \ \ \xi \in \mathbb{R}.
\end{equation}
\end{lemma}

In the proof of \cref{thm:pbp_corr}, we will make extensive use of two auxiliary functions. 

\begin{definition}[Auxiliary functions]
Let $\mathcal{P}$ and $\mathcal{T}$ be defined as follows:
\begin{equation}\label{eq:P}
\mathcal{P}(\xi) \Equaldef \xi - \eta(\xi), \ \ \xi\in \mathbb{R}
\end{equation}
and
\begin{equation}\label{eq:T}
\mathcal{T}(\xi) \Equaldef \xi + \eta(\xi), \ \ \xi\in \mathbb{R}.
\end{equation}
\end{definition}
The fact that $\mathcal{P}$ and $\mathcal{T}$ possess certain monotonicity properties is of paramount importance in the proof of \cref{thm:pbp_corr}.
\begin{lemma}[Monotonicity of $\mathcal{P}$ and $\mathcal{T}$]\label{lem:monotonicity}
Let $\xi_1,\xi_2 \in \mathbb{R}$. For all $\xi_1$ and $\xi_2$ such that $\xi_1 \leq \xi_2$, then:
\begin{IEEEeqnarray}{rCl}
\mathcal{P}(\xi_1) & \leq & \mathcal{P}(\xi_2)\\
\mathcal{T}(\xi_1) & \leq & \mathcal{T}(\xi_2).
\end{IEEEeqnarray}
\end{lemma}

\begin{IEEEproof}
See Appendix A.
\end{IEEEproof}



We are now equipped to prove \cref{thm:pbp_corr}.

\begin{IEEEproof}[Proof of \Cref{thm:pbp_corr}]
Assuming that $\mathcal{U}=\mathcal{U}^{\max}$, due to the symmetry of the pdf, \cref{lem:estimator_identity} implies that the optimal estimator $\mathcal{E}^{\star}_{\mathcal{U}^{\max}}$ is characterized by a single representation function $\eta(\xi)$ in \cref{eq:nonlinear} as follows:
\begin{equation}
\mathcal{E}^{\star}_{\mathcal{U}^{\max}}(1,\xi) = \begin{bmatrix} \xi \\ \eta(\xi)\end{bmatrix} \ \ \text{and} \ \ \mathcal{E}^{\star}_{\mathcal{U}^{\max}}(2,\xi) = \begin{bmatrix} \eta(\xi) \\ \xi\end{bmatrix}.
\end{equation}
Therefore, 
\begin{equation}
\mathcal{E}^{\star}_{\mathcal{U}^{\max}} = \mathcal{E}^{\mathrm{soft}}.
\end{equation}

We will show that this choice of estimation policy implies, via \cref{lem:scheduler}, the optimality of the max-scheduling policy. Define the function $\mathcal{H}:\mathbb{R}^2\rightarrow \mathbb{R}$ such that:
\begin{equation}
\mathcal{H}(x) \Equaldef \big(x_2-\eta(x_1))^2 - (x_1-\eta(x_2)\big)^2.
\end{equation} 
The function above can be rewritten using the two auxiliary functions $\mathcal{P}$ and $\mathcal{T}$ from \cref{eq:P,eq:T} as follows:
\begin{equation}
\mathcal{H}(x) = \big[ \mathcal{T}(x_2) - \mathcal{T}(x_1) \big] \times \big[\mathcal{P}(x_2)+\mathcal{P}(x_1)\big].
\end{equation} 
\cref{lem:scheduler} implies the optimal scheduling policy given by:
\begin{equation}
\mathcal{U}^{\star}_{\mathcal{E}^{\star}_{\mathcal{U}^{\max}}}(x) = \begin{cases} 1 &  \text{if} \ \ \mathcal{H}(x) \leq 0  \\ 2 & \text{otherwise.} \end{cases}
\end{equation}
Partition of $\mathbb{R}^2$ into eight subsets $\big\{\mathbb{A}_1,\cdots,\mathbb{A}_8\big\}$ depicted in \cref{fig:partition}. Let $x\in \mathbb{A}_1$, which is characterized by $x_1\geq 0$, $x_2\geq 0$ and $x_1\geq x_2$. \cref{lem:monotonicity} implies that:
\begin{IEEEeqnarray}{rCl}
\mathcal{P}(x_1) &\geq& \mathcal{P}(0) = 0\\
\mathcal{P}(x_2) &\geq& \mathcal{P}(0) = 0.
\end{IEEEeqnarray}
Therefore,
\begin{equation}\label{eq:ineq1}
\mathcal{P}(x_1) + \mathcal{P}(x_2) \geq 0.
\end{equation}
Since $x_2\geq x_1$, \cref{lem:monotonicity} also implies that:
\begin{equation}\label{eq:ineq2}
\mathcal{T}(x_2) - \mathcal{T}(x_1) \geq 0.
\end{equation}
Together, \cref{eq:ineq1,eq:ineq2} imply:
\begin{equation}
\mathcal{H}(x) \geq 0.
\end{equation}

\begin{figure}[!b]
    \begin{center}
    \includegraphics[width=0.5\textwidth]{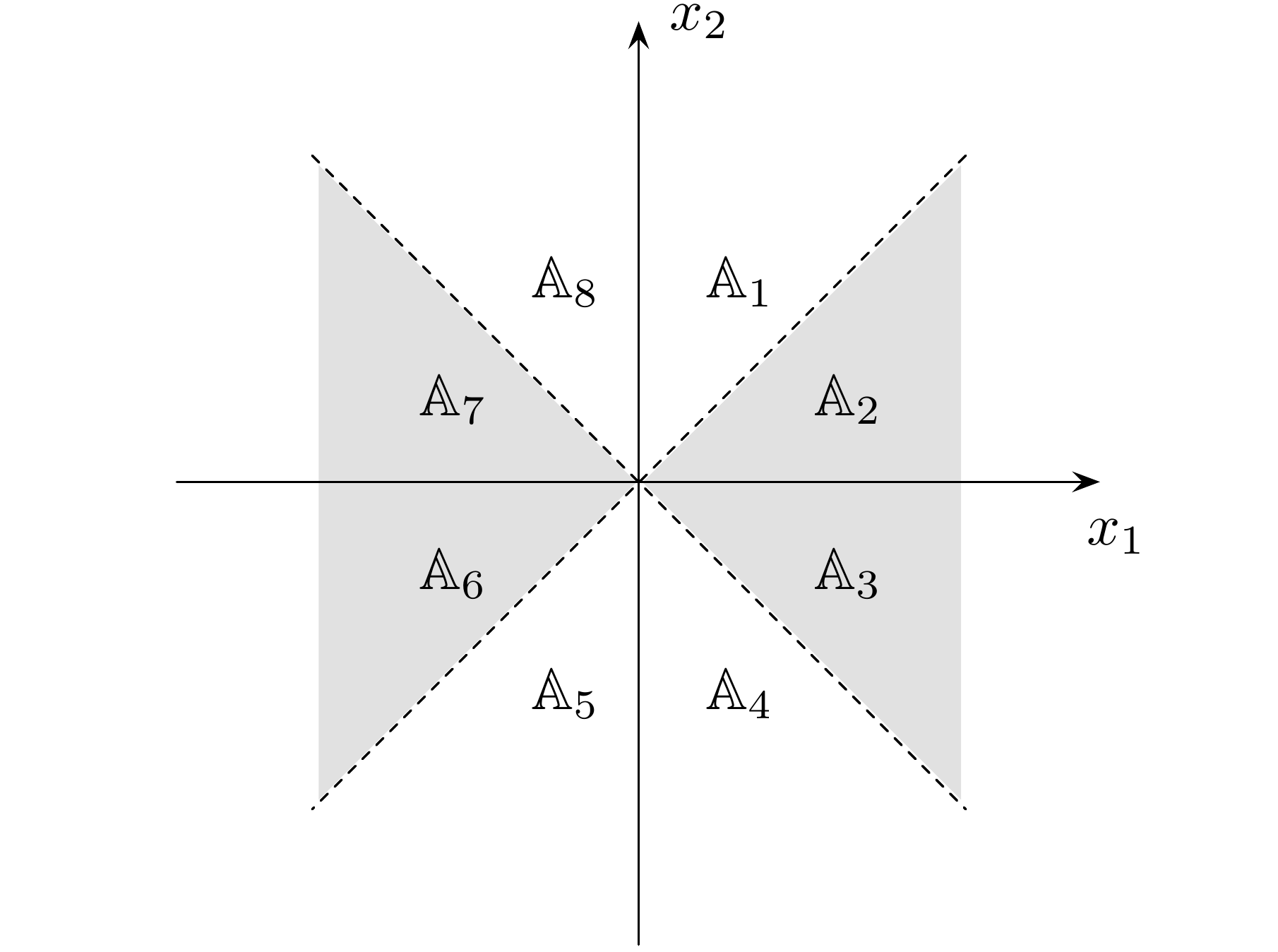}
\end{center}
\caption{Partition of the observation space used in the proof of \Cref{thm:pbp_corr}.}
\label{fig:partition}
\end{figure}

Similarly, if $x\in \mathbb{A}_2$, we have:
\begin{equation}
\mathcal{P}(x_1) + \mathcal{P}(x_2) \geq 0.
\end{equation}
On the other hand:
\begin{equation}
\mathcal{T}(x_2) - \mathcal{T}(x_1) \leq 0.
\end{equation}
Therefore,
\begin{equation}
\mathcal{H}(x) \leq 0.
\end{equation}

Proceeding in a similar way, making use of the \cref{lem:odd,lem:monotonicity}, we can cover all the eight regions. Thereby, showing that for each of the eight regions all the points either satisfy $\mathcal{H}(x)\geq 0$ or $\mathcal{H}(x)\leq 0$. Moreover, if $x\in \mathbb{A}_1 \cup \mathbb{A}_4 \cup \mathbb{A}_5 \cup \mathbb{A}_8$ then $\mathcal{H}(x) \geq 0$ and if $x\in \mathbb{A}_2 \cup \mathbb{A}_3 \cup \mathbb{A}_6 \cup \mathbb{A}_7$ then $\mathcal{H}(x) \geq 0$. Thus, equivalently showing that:
\begin{equation}
\mathcal{U}^{\star}_{\mathcal{E}^{\mathrm{soft}}} = \mathcal{U}^{\max}.
\end{equation}

\end{IEEEproof}

\section{The decorrelating transformation approach}



\begin{figure*}[t!]
\centering
\includegraphics[scale=0.4]{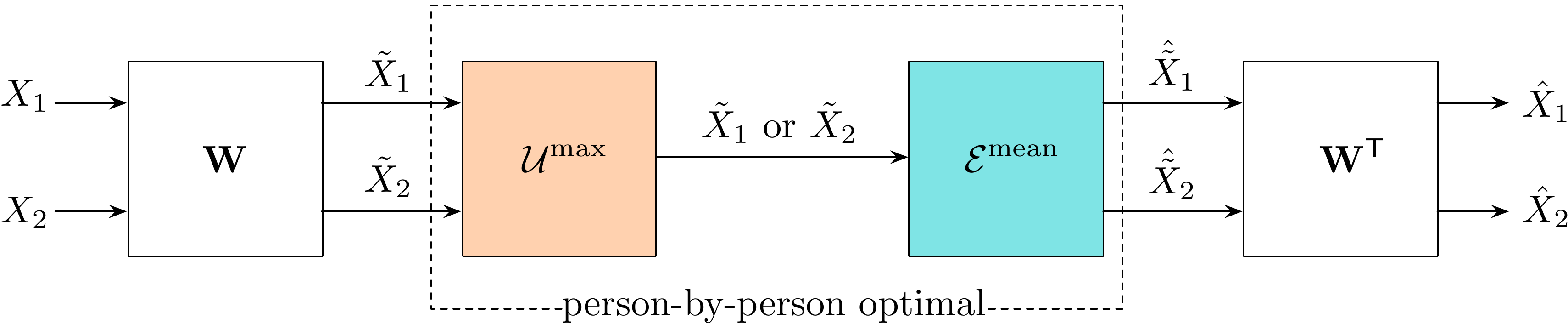}
\caption{System architecture for the arbitrarily correlated case. The pre-processing block implements a decorrelating linear transformation $\mathbf{W}$ obtained from the eigendecomposition of the covariance matrix $\mathbf{\Sigma}$. The post-processing block implements the inverse of the decorrelating transformation $\mathbf{W}^{\mathsf{T}}$.}
\label{fig:whitening}
\end{figure*}

In this section, we propose a person-by-person optimal solution to the scheduling of two arbitrarily correlated Gaussian sources by using pre- and post-processing blocks on the observations and the estimates. The idea is to ``decorrelate'' the two observations using an invertible linear transformation, use the max-scheduling/mean-estimation policy on the transformed random variables, and then ``correlate'' the estimates using the inverse transformation. This strategy is depicted in the block diagram of \cref{fig:whitening}.

Consider the eigendecomposition of symmetric positive definite covariance matrix $\mathbf{\Sigma}$:
\begin{equation}
\mathbf{\Sigma} = \mathbf{W} \mathbf{\Lambda} \mathbf{W}^{\mathsf{T}},
\end{equation}
where $\mathbf{W}\mathbf{W}^{\mathsf{T}}=\mathbf{I}$, and $\mathbf{\Lambda}$ is a diagonal matrix. Using the matrix $\mathbf{W}$, define the following scheduling and estimation policies:
\begin{equation}
\mathcal{U}^{\mathrm{dec}}(x) \Equaldef \mathcal{U}^{\max}\big(\mathbf{W}x \big), \ \ x\in\mathbb{R}^2
\end{equation}
and
\begin{equation}
\mathcal{E}^{\mathrm{dec}}(i,\xi) \Equaldef \mathbf{W}^{\mathsf{T}}\mathcal{E}^{\mean}(i,\xi), \ \ i \in \{1,2\}, \ \ \ \xi \in \mathbb{R}.
\end{equation}

\begin{theorem}
Let $X\sim\mathcal{N}(\mathbf{0},\mathbf{\Sigma})$, where $\mathbf{\Sigma}$ is a symmetric positive definite covariance matrix. The pair $(\mathcal{U}^{\mathrm{dec}},\mathcal{E}^{\mathrm{dec}})$ is a person-by-person optimal solution to \Cref{prob:main}.
\end{theorem}

\begin{IEEEproof}
Let $\mathbf{W}$ be computed from the eigendecomposition of $\mathbf{\Sigma}$, and denote
\begin{equation}
\mathbf{W} = \begin{bmatrix}
w_{11} & w_{12} \\ w_{21} & w_{22}
\end{bmatrix}.
\end{equation}
Let $\tilde{x}\in \mathbb{R}^2$ be defined as $\tilde{x}\Equaldef \mathbf{W}x$. Assuming that the estimator uses policy $\mathcal{E}^{\mathrm{dec}}$, then the optimal scheduling decision rule is to transmit $\tilde{x}_1$ if:
\begin{equation}
(x_1-w_{11}\tilde{x}_1)^2 + (x_2-w_{12}\tilde{x}_1)^2  \leq  (x_1-w_{21}\tilde{x}_2)^2 + (x_2-w_{22}\tilde{x}_2)^2.
\end{equation}
Recalling that $x=\mathbf{W}^{\mathsf{T}}\tilde{x}$, we have:
\begin{equation}
(w_{21}^2+w_{22}^2)\tilde{x}_2^2  \leq (w_{11}^2+w_{12}^2)\tilde{x}_1^2 \end{equation}
Since $\mathbf{W}$ is a unitary matrix, the inequality above is equivalent to:
\begin{equation}
|\tilde{x}_2|  \leq |\tilde{x}_1|.
\end{equation}
Therefore, 
\begin{equation}
\mathcal{U}^{\star}_{\mathcal{E}^{\mathrm{dec}}} = \mathcal{U}^{\mathrm{dec}}.
\end{equation}

Conversely, assume that the scheduler uses $\mathcal{U}^{\mathrm{dec}}$. Let $i,j\in \{1,2\}$ such that $i\neq j$. Then,
\begin{equation}
\mathcal{E}^{\star}_{\mathcal{U}^{\mathrm{dec}}}(i,\tilde{x}_i) = \mathbf{W}^{\mathsf{T}}\mathbf{E}\big[\tilde{X} \mid Y=
(i,\tilde{x}_i) \big],
\end{equation}
where 
\begin{equation}
\tilde{X} \Equaldef \mathbf{W}X.
\end{equation}
Computing the expectation above gives:
\begin{equation}
\mathbf{E}\big[ \tilde{X}_i \mid Y=(i,\tilde{x}_i)\big] = \tilde{x}_i,
\end{equation}
and, for $j\neq i$, we have:
\begin{equation}
\mathbf{E}\big[ \tilde{X}_j \mid Y=(i,\tilde{x}_i)\big]  = \frac{\int_{-|\tilde{x}_i|}^{|\tilde{x}_i|}\tilde{x}_jf_{\tilde{X}_j|\tilde{X}_i=\tilde{x}_i}(\tilde{x}_j)d\tilde{x}_j}{\int_{-|\tilde{x}_i|}^{|\tilde{x}_i|}f_{\tilde{X}_j|\tilde{X}_i=\tilde{x}_i}(\tilde{x}_j)d\tilde{x}_j}.
\end{equation}
Since $\tilde{X}_i \perp \!\!\! \perp \tilde{X}_j$, and $f_{\tilde{X}_j}$ is an even function, we have:
\begin{equation}
\mathbf{E}\big[ \tilde{X}_j \mid Y=(i,\tilde{x}_i)\big]  = 0, \ \ \tilde{x}_i \in \mathbb{R}.
\end{equation}
Therefore, 
\begin{equation}
\mathcal{E}^{\star}_{\mathcal{U}^{\mathrm{dec}}} = \mathcal{E}^{\mathrm{dec}}.
\end{equation}

\end{IEEEproof}




\begin{remark}
Despite the fact that $(\mathcal{U}^{\mathrm{dec}},\mathcal{E}^{\mathrm{dec}})$ is person-by-person optimal for Problem 1, we will show later that this is a suboptimal solution in general. For example, for a symmetric correlated source, the pair $(\mathcal{U}^{\mathrm{max}},\mathcal{E}^{\mathrm{soft}})$ yields a smaller cost, albeit the difference in performance is not large. We conjecture that, in general, the globally optimal estimation policy is nonlinear. However, Theorem 3 is a useful result because it leads to person-by-person optimal policies for Gaussian sources of arbitrary dimension as we will formally state in Section VIII.
\end{remark}

\section{Linear Minimum Mean Squared Error Estimators}

Up to this point, we have obtained person-by-person optimal solutions to \cref{prob:main}, which is defined over infinite dimensional policy spaces. In this section, we will consider the design of jointly optimal scheduling and estimation policies when the estimation policies are constrained to belong to the parametrizable class of \textit{piecewise linear} estimation policies.

\begin{definition}[Class of admissible piecewise linear estimation policies]
Let $a\in \mathbb{R}^2$. An admissible estimation policy $\mathcal{E}^{\mathrm{linear}}_a \in \mathbb{E}$ is piecewise linear if it has the following structure:
\begin{equation}
\mathcal{E}^{\mathrm{linear}}_a(i,\xi) = \begin{cases} \big[\ \xi  \ \ \  a_2\xi \ \big]^\mathsf{T} & \text{if} \ \ i=1 \\
\big[ \ a_1\xi \ \ \ \xi \ \big]^{\mathsf{T}} & \text{if} \ \ i=2.
\end{cases}
\end{equation}
The set of all admissible piecewise linear estimation policies is denoted by $\mathbb{E}^{\mathrm{linear}}$.
\end{definition}



\subsection{LMMSE estimation of symmetric correlated sources}

Within the class of piecewise linear estimators, \cref{prob:main} admits a unique solution when the sources are symmetric. Before stating this result in Theorem 4, we show that the search for LMMSE estimators can be performed by solving a finite dimensional optimization problem. 

\begin{proposition}
Consider \cref{prob:main} with the additional constraint that $\mathcal{E}\in \mathbb{E}^{\mathrm{linear}}$. Then, the problem is equivalent to the finite dimensional nonconvex optimization problem below:
\begin{equation}\label{eq:nlin_int_prog}
\begin{aligned}
& \underset{a \in \mathbb{R}^2}{\mathrm{minimize}}
& & \mathcal{J}_q(a)
\end{aligned}
\end{equation}
where the objective function $\mathcal{J}_q: \mathbb{R}^2 \rightarrow \mathbb{R}$ is defined as:
\begin{equation}\label{eq:finite_dim_prob}
\mathcal{J}_q (a) \Equaldef \mathbf{E}\Big[\min\Big\{\big(X_1-a_1X_2\big)^2, \big(X_2-a_2X_1\big)^2 \Big\}\Big].
\end{equation}
\end{proposition}

\begin{IEEEproof}
Recalling the cost functional:
\begin{equation}
\mathcal{J}(\mathcal{U},\mathcal{E}) = \mathbf{E}\big[(X_1-\hat{X}_1)^2+(X_2-\hat{X}_2)^2\big].
\end{equation}
If $\mathcal{E} \in \mathbb{E}^{\mathrm{linear}}$, then the cost can be rewritten in integral form as:
\begin{equation}
\mathcal{J}(\mathcal{U},\mathcal{E}) = \int_{\mathbb{R}^2} \big(x_1-a_1x_2\big)^2\mathbf{1}(\mathcal{U}(x)=2)f_X(x)dx 
+\int_{\mathbb{R}^2} \big(x_2-a_2x_1\big)^2\mathbf{1}(\mathcal{U}(x)=1)f_X(x)dx 
\end{equation}
For arbitrarily fixed constants $a_1,a_2 \in \mathbb{R}$, the optimal scheduling policy $\mathcal{U}_{a}^{\star}$ is given by:
\begin{equation}
\mathcal{U}_{a}^{\star}(x) \Equaldef \begin{cases} 1 & \text{if}  \ \ \big(x_2 - a_2x_1\big)^2 \leq \big(x_1 - a_1x_2\big)^2 \\
2 & \text{otherwise.}
\end{cases}
\end{equation}
Therefore, we may, without loss of optimality, define a new cost solely in terms of $a\in \mathbb{R}^2$:
\begin{equation}
\mathcal{J}_q(a) \Equaldef \mathcal{J}(\mathcal{U}^{\star}_a,\mathcal{E}),
\end{equation}
which is equal to the expression in \cref{eq:finite_dim_prob}.
\end{IEEEproof}



\begin{theorem}
Consider two symmetric correlated Gaussian sources with variance $\sigma^2$ and correlation coefficient $\rho$. Constraining the estimator to belong to the class of piecewise linear functions, the policy pair $(\mathcal{U}^{\max},\mathcal{E}^{\mathrm{linear}}_{a^{\star}})$ is globally optimal for \cref{prob:main}, where:
\begin{equation}
a^{\star} = \frac{\rho \cdot \sigma^2}{2\cdot\int_{\mathbb{R}^2}x_1^2 \mathbf{1}(|x_1|\geq |x_2|)f_X(x)dx}.
\end{equation}
\end{theorem}

\begin{IEEEproof}
See Appendix B.
\end{IEEEproof}

\begin{figure}[!t]
    \begin{center}
    \includegraphics[width=0.5\textwidth]{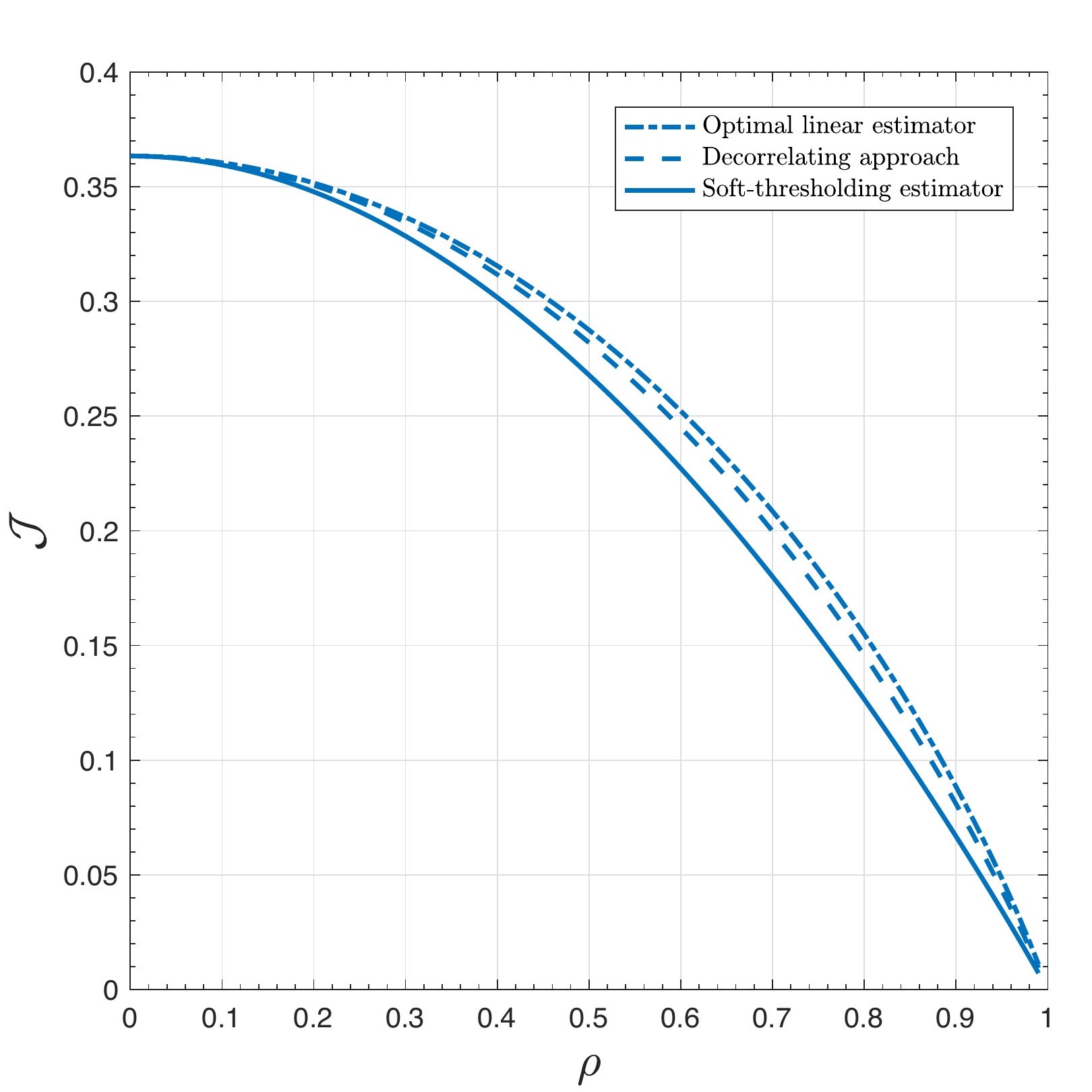}
\end{center}
\caption{Performance of three different scheduling and estimation schemes for symmetric correlated Gaussian sources with variance $\sigma^2=1$.}
\label{fig:performance_corr}
\end{figure}

\begin{remark}
The performance of the scheduling/estimation schemes of Sections V, VI and VII-A are displayed in \cref{fig:performance_corr} for symmetric correlated Gaussian sources with variance $\sigma^2=1$. The system implementing max-scheduling and nonlinear soft-thresholding estimation of Theorem 2 has the best performance. We conjecture that this is indeed the globally optimal performance in this case. The performance of the decorrelating transformation approach followed by max-scheduling and mean-estimation of Theorem 3 has the second best performance, and as we can see, it shows that a person-by-person optimal solution is not necessarily optimal. Finally, the worst performance is of max-scheduling followed by the optimal linear estimator of Theorem 4. Despite being suboptimal, this is a globally optimal solution among the class of all possible piecewise linear estimators. Therefore, we can trustfully state that this solution cannot be improved upon, whereas the other two strategies do not share this feature.
\end{remark}

\subsection{Optimization via the Convex-Concave Procedure}

Notice that the equivalent optimization problem stated in Proposition 1, although finite dimensional, it is still a non-convex stochastic program. Unlike the symmetric case, a closed form solution to this problem for the general case is {}not known. However, in our subsequent analysis we will decompose the cost into a difference of convex functions and derive an efficient numerical optimization algorithm to compute locally optimal solutions using the so-called \textit{Convex-Concave Procedure} \cite{Yuille:2003}.

Define the functions $\mathcal{F},\mathcal{G}$: $\mathbb{R}^2\rightarrow \mathbb{R}$ such that:
\begin{equation}
\mathcal{F}(a) \Equaldef (1+a_2^2)\sigma_1^2 + (1+a_1^2)\sigma_2^2  -2 \rho \sigma_1 \sigma_2 (a_1 +a_2) 
\end{equation}
and
\begin{equation}\label{eq:G}
\mathcal{G}(a) \Equaldef \mathbf{E} \Big[\max\Big\{\big(X_1-a_1X_2\big)^2, \big(X_2-a_2X_1\big)^2 \Big\} \Big].
\end{equation}
The cost function in \cref{eq:finite_dim_prob} can be expressed as a difference of convex functions as follows:
\begin{equation}
\mathcal{J}_q(a) = \mathcal{F}(a) - \mathcal{G}(a).
\end{equation}

The convex-concave procedure (CCP) for minimizing $\mathcal{J}_q$ is given by the following algorithm:
\begin{equation}\label{eq:CCP}
 a^{(k+1)} = \arg\min_{a\in \mathbb{R}^2} \Big\{ \mathcal{F}(a) - \mathcal{G}_{\mathrm{affine}}(a;a^{(k)})\Big\},
\end{equation}
where
\begin{equation}
\mathcal{G}_{\mathrm{affine}}(a;a^{(k)}) \Equaldef \mathcal{G}(a^{(k)}) + g(a^{(k)})^{\mathsf{T}} ( a - a^{(k)} )
\end{equation}
and $g(a)$ is a subgradient of $\mathcal{G}(a)$. We solve the optimization problem in \cref{eq:CCP}, by using the first order optimality condition:
\begin{equation}
\nabla\mathcal{F}(a^{\star}) - g(a^{(k)}) = \mathbf{0}.
\end{equation}

Since the function $\mathcal{F}$ is differentiable in both of its arguments, its gradient can be explicitly computed:
\begin{equation}
\nabla \mathcal{F} (a) = \begin{bmatrix}
2a_1\sigma_2^2 -2\rho\sigma_1\sigma_2 \\
2a_2\sigma_1^2 -2\rho\sigma_1\sigma_2
\end{bmatrix},
\end{equation}
which leads to the  following dynamical system:
\begin{equation}
\begin{bmatrix} a_1^{(k+1)} \\ a_2^{(k+1)}  \end{bmatrix} =
\begin{bmatrix} 
\frac{1}{2\sigma_2^{2}} & 0 \\
0 & \frac{1}{2\sigma_1^{2}}  \\
  \end{bmatrix} 
 g(a^{(k)})  %
  + \begin{bmatrix} \rho\frac{\sigma_1}{\sigma_2} \\ \rho\frac{\sigma_2}{\sigma_1}  \end{bmatrix}.
\end{equation}

The sequence $\{a_k\}_{k=1}^{\infty}$ defined by the system above always converges to a critical point of $\mathcal{J}_q$ \cite{Lipp:2016}. In order to compute a subgradient $g(a)$, we use the rules of (weak) subgradient calculus \cite{Boyd:2018}.



\begin{proposition}
The map $g: \mathbb{R}^2 \rightarrow \mathbb{R}^2$ defined as:
\begin{equation}
g(a) \Equaldef -2\cdot\mathbf{E}\begin{bmatrix}
(X_1-a_1X_2)\cdot X_2 \cdot \mathbf{1}\big(|X_1-a_1X_2|\geq |X_2-a_2X_1|\big)\\(X_2-a_2X_1)\cdot X_1 \cdot \mathbf{1}\big(|X_1-a_1X_2| < |X_2-a_2X_1|\big)
\end{bmatrix}
\end{equation}
\normalsize
is a subgradient of $\mathcal{G}(a)$ defined in \cref{eq:G}.
\end{proposition}

\begin{IEEEproof}
Let the function $\mathcal{G}(a;x)$ be defined as:
\begin{equation}
\mathcal{G}(a;x) \Equaldef \max\big\{\mathcal{G}_1(a;x),\mathcal{G}_2(a;x)\big\},
\end{equation}
where
\begin{equation}
\mathcal{G}_1(a;x) \Equaldef (x_1-a_1x_2)^2
\end{equation}
and
\begin{equation}
\mathcal{G}_2(a;x) \Equaldef (x_2-a_2x_1)^2.
\end{equation}
Therefore, 
\begin{equation}
\mathcal{G}(a) = \mathbf{E}\big[\mathcal{G}(a;X)\big].
\end{equation}

In order to construct a subgradient of $\mathcal{G}(a)$, we first find a subgradient $g(a;x)$ of $\mathcal{G}(a;x)$ and take its expectation with respect to $x$.

At the points where $\mathcal{G}_1(a;x) > \mathcal{G}_2(a;x)$, then
\begin{equation}
g(a;x) \Equaldef
\nabla \mathcal{G}_1(a;x) = \begin{bmatrix} -2(x_1-a_1x_2)x_2 \\ 0\end{bmatrix}.
\end{equation}
Similarly, at the points where $\mathcal{G}_1(a;x) < \mathcal{G}_2(a;x)$, then
\begin{equation}
g(a;x) \Equaldef
\nabla \mathcal{G}_2(a;x) = \begin{bmatrix} 0 \\ -2(x_2-a_2x_1)x_1\end{bmatrix}.
\end{equation}
When $\mathcal{G}_1(a;x) = \mathcal{G}_2(a;x)$, either one of the gradients above can be chosen as a subgradient of $\mathcal{G}(a;x)$. The following choice is a valid subgradient of $\mathcal{G}(a;x)$:
\begin{equation}
g(a;x) = \nabla \mathcal{G}_1(a;x)\mathbf{1}\big(\mathcal{G}_1(a;x)\geq \mathcal{G}_2(a;x)\big) \\ + \nabla \mathcal{G}_2(a;x)\mathbf{1}\big(\mathcal{G}_1(a;x) < \mathcal{G}_2(a;x)\big).
\end{equation}
Finally, we let
\begin{equation}
g(a) \Equaldef \mathbf{E}\big[g(a;X)\big],
\end{equation}
which is a subgradient of $\mathcal{G}(a)$.
\end{IEEEproof}

\begin{figure}[!t]
    \begin{center}
    \includegraphics[width=0.5\textwidth]{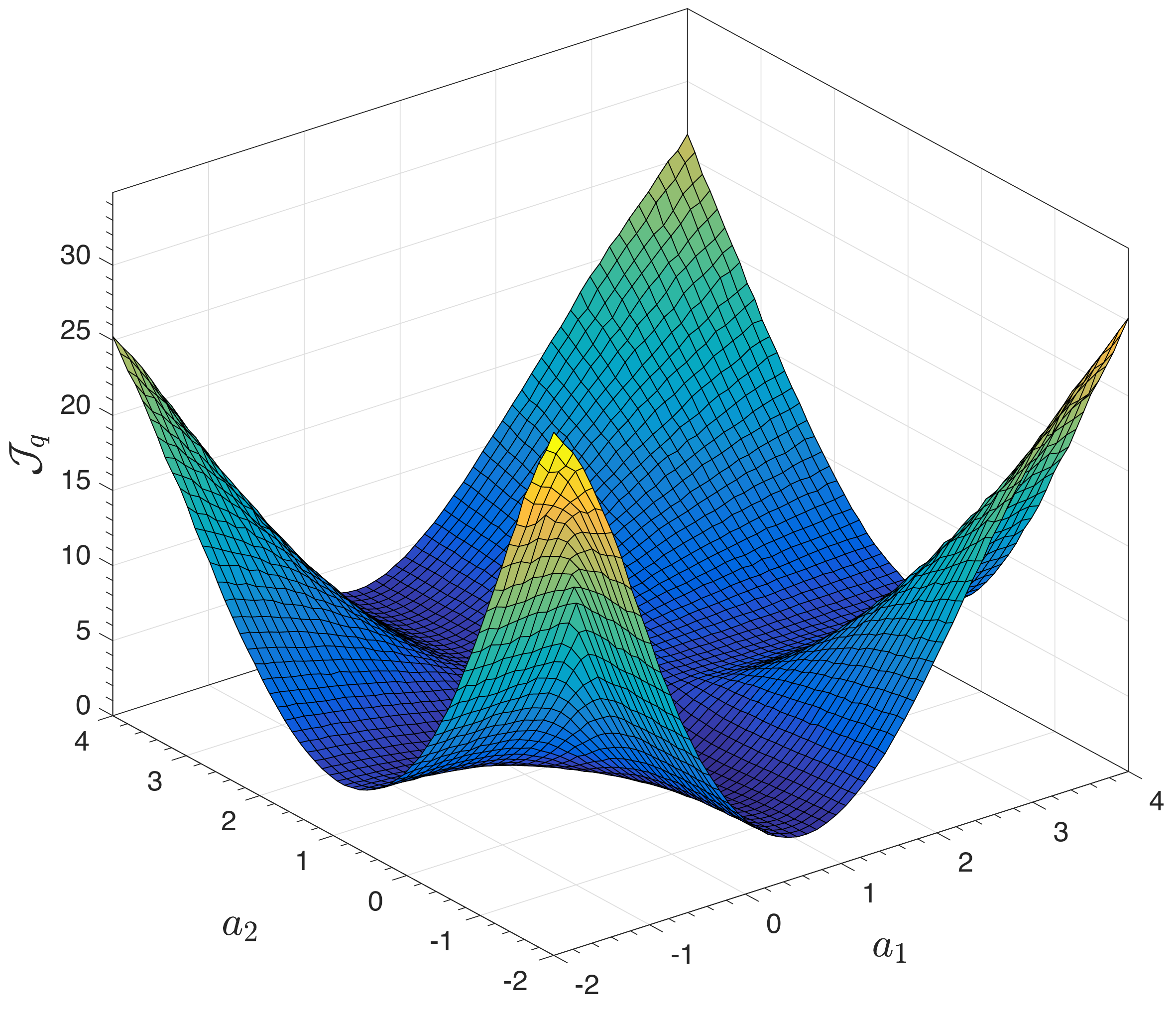}
\end{center}
\caption{Cost function $\mathcal{J}_q$ for two correlated Gaussian sources with parameters $\sigma_1^2=5$, $\sigma_2^2=7$ and $\rho=0.6$. A single global minimum exists, but a formal proof is difficult to obtain due to the lack of convexity.}
\label{fig:Cost_ODSS}
\end{figure}

\begin{remark}
In principle, the CCP algorithm above does not guarantee that the solutions found through the algorithm are globally optimal. However, in the case considered in this paper where two scalar Gaussian sources are being scheduled, the finite dimensional cost function can be visualized (as in \cref{fig:Cost_ODSS}) and the global optimality of the solutions can be empirically verified. \Cref{tab:CCP} shows the solutions found by the CCP algorithm 
for sources with variances $\sigma_1^2=5$ and $\sigma^2_2=7$ and several values of the correlation coefficient $\rho$. For all the entries in \Cref{tab:CCP} the solutions were verified to be globally optimal. One advantage of this numerical scheme is that it can be used for jointly designing schedulers and piecewise linear estimators for any pair of sources, \underline{regardless of their joint distribution}.
\end{remark}

\begin{table}[b!]
\centering
\caption{Optimal cost for scheduling policies induced by piecewise linear estimation policies for Gaussian sources with $\sigma^2_1=5$ and $\sigma_2^2=7$.}
\begin{tabular}{cccc}
$\rho$ & $\mathcal{J}_q^{\star}$ & $a_1^{\star}$ & $a_2^{\star}$ \\ \hline \hline
$0  $ & $2.1271$ & $0.0007$ & $0.0012$ \\
$0.1$ & $2.1099$ & $0.0552$ & $0.0678$ \\
$0.2$ & $2.0579$ & $0.1131$ & $0.1330$ \\
$0.3$ & $1.9704$ & $0.1709$ & $0.2023$ \\
$0.4$ & $1.8457$ & $0.2292$ & $0.2772$ \\
$0.5$ & $1.6815$ & $0.2936$ & $0.3513$ \\
$0.6$ & $1.4741$ & $0.3612$ & $0.4345$ \\
$0.7$ & $1.2179$ & $0.4336$ & $0.5314$ \\
$0.8$ & $0.9038$ & $0.5189$ & $0.6426$ \\
$0.9$ & $0.5149$ & $0.6255$ & $0.7897$
 \\ \hline \hline
\end{tabular}
\label{tab:CCP}
\end{table}

\section{Extensions}

\Cref{thm:pbp} can be extended to any number of sensors observing independent zero mean Gaussian random variables. This is a significant generalization of the two sensor case considered in Section IV. Let $x\in \mathbb{R}^n$ and consider the following generalization to the max-scheduling and mean-estimation strategies for $n\geq 2$:
\begin{equation}\label{eq:Umax}
\mathcal{U}^{\max}(x) \Equaldef \arg \max_{i\in\{1,\cdots,n\}} |x_i|
\end{equation}
and
\begin{equation}\label{eq:Emean}
\mathcal{E}^{\mathrm{mean}}(i,\xi)\Equaldef \xi\cdot\mathbf{e}_i,
\end{equation}
where $\mathbf{e}_i$ is the $i$-th standard basis vector in $\mathbb{R}^n$, $i\in\{1,2,\cdots,n\}$ and $\xi \in \mathbb{R}$.

\begin{theorem}
 If $X\sim\mathcal{N}\big(\mathbf{0},\mathbf{diag}(\sigma_1^2, \sigma_2^2, \cdots,\sigma_n^2)\big)$, then $(\mathcal{U}^{\max},\mathcal{E}^{\mean})$ is a person-by-person optimal solution to Problem 1.
\end{theorem}

At this point it is unclear if Theorem 2 can also be generalized to an arbitrary number of sensors with  correlation structures that display some kind of symmetry, and it is a topic for future research. However, Theorem 5 is useful for the scheduling of $n$ sensors with arbitrary correlation matrix provided that we use the pre- and post-processing using the decorrelating transformation derived from its eigendecomposition in the same spirit of Section VI. We present the generalization of Theorem 3 to an arbitrarily $n$-dimensional correlated Gaussian source.

\begin{theorem}
 Let $X\sim\mathcal{N}\big(\mathbf{0},\mathbf{\Sigma})$ and the unitary matrix $\mathbf{W}$ be obtained from the eigendecomposition of the covariance matrix as $\mathbf{\Sigma} = \mathbf{W}\mathbf{\Lambda}\mathbf{W}^{\mathsf{T}}$. Define 
 \begin{equation}
\mathcal{U}^{\mathrm{dec}} (x) \Equaldef \mathcal{U}^{\mathrm{max}} (\mathbf{W}x)
 \end{equation}
 and
  \begin{equation}
\mathcal{E}^{\mathrm{dec}} (i,\xi) \Equaldef \mathbf{W}^{\mathsf{T}}\mathcal{E}^{\mathrm{mean}} (i,\xi),
 \end{equation}
 where $\mathcal{U}^{\max}$ and $\mathcal{E}^{\mean}$ are given by \cref{eq:Umax,eq:Emean}.
The pair $(\mathcal{U}^{\mathrm{dec}},\mathcal{E}^{\mathrm{dec}})$ is a person-by-person optimal solution to Problem 1.
 
\end{theorem}

\section{Conclusions}
We have presented a new approach to sensor scheduling where a centralized agent observes the realization of a bivariate Gaussian source and chooses a single component to be transmitted to a remote estimator. The motivation for this problem comes from constraints in networked control and estimation, where a single packet can be reliably transmitted over a communication link. This problem can also be viewed as a \textit{data-driven} dimensionality reduction problem. Similarly to the ``Witsenhausen's counter-example'', the design of globally optimal scheduling and estimation policies is elusive due to the lack of convexity of the overall optimization problem caused by signaling. However, we can prove the person-by-person optimality of a pair of policies in two important particular cases: independent observations; and symmetrically correlated observations. In both cases the optimal (in the person-by-person sense) scheduling policy consists of selecting the measurement with the largest magnitude to be transmitted. Interestingly, this scheduling scheme provides the receiver with upper and lower bounds which can be used for estimating the non-transmitted measurement. In the independent case, the person-by-person optimality result can be extended to any number of sensors. We also showed how to use the first person-by-person result to obtain suboptimal policies for the general correlated Gaussian case, where a pre-processing linear decorrelating transformation followed by a max-scheduler is used. On the receiver end, the mean-estimation policy is followed by the inverse transformation. Finally, we have considered the joint design of scheduling and estimation policies for a bivariate Gaussian source when the estimator is constrained to the class of piecewise linear estimators. In this case, we obtained globally optimal solutions to the non-convex optimization problem by using the Convex-Concave Procedure.

\underline{Opportunities for future work}: This work aims to initiate an entire class of problems in sensor scheduling, which we refer to as \textit{observation-driven sensor scheduling}. One important topic for future investigation is to prove the conjecture that the pair $(\mathcal{U}^{\max},\mathcal{E}^{\mean})$ is globally optimal for the two main cases in the first part of the paper. We believe that the proof of global optimality will involve results from information theory, such as the data-processing inequality and rate-distortion function for Gaussian sources as in \cite{Choudhuri:2012}. A similar approach was used in \cite{Basar:2008} to prove the joint optimality of linear transmission and estimation policies for the ``Gaussian Test Channel''. A second topic for future work is to bound the performance of the separated design using the decorrelating transformation for an arbitrary number of sensors, in order to obtain a performance guarantee. Also, extending the person-by-person optimality result for a Gaussian source with a correlated structure under suitable symmetry structures in the covariance matrix. Finally, to investigate how this theory would apply for more general joint distributions, and possibly accounting for situations where the joint distribution is unknown or needs to be estimated from data.       


\appendices

\section{Monotonicity of $\mathcal{P}$ and $\mathcal{T}$}

The proof of \Cref{thm:pbp_corr} requires that the functions $\mathcal{P}$ and $\mathcal{T}$ defined in \cref{eq:P,eq:T} are monotone increasing. Here we present a proof of \Cref{lem:monotonicity}.

\begin{IEEEproof}[Proof of \Cref{lem:monotonicity}]
Consider the function $\mathcal{P}$. Recall that 
\begin{equation}
\mathcal{P}(\alpha) \Equaldef \alpha - \eta(\alpha),
\end{equation}
where $\eta$ is defined in \cref{eq:nonlinear}. The function $\mathcal{P}$ can be alternatively expressed as:
\begin{equation}
\mathcal{P}(\alpha) = \mathbf{E} \Big[\alpha-X \ \Big|\ -|\alpha| \leq X \leq |\alpha| \Big],
\end{equation}
where $X \sim \mathcal{N}\big(\rho\alpha,\sigma^2(1-\rho^2)\big)$.
Since $\mathcal{P}$ is an odd function, we can constrain our analysis to $\alpha\geq 0$, without loss of generality. Therefore, we assume that:
\begin{equation}
\mathcal{P}(\alpha) = \mathbf{E} \Big[\alpha-X \ \Big|\ -\alpha \leq X \leq \alpha \Big].
\end{equation}
Notice that, when conditioned on $\{-\alpha\leq X \leq \alpha\}$, the following inequality holds:
\begin{equation}
\alpha-X \geq 0 \ \ \mathrm{a.s.} 
\end{equation}
Therefore, we can rewrite $\mathcal{P}$ as: 
\begin{equation}
\mathcal{P}(\alpha) = \int_{0}^{\infty} \Pr\Big( \alpha-X \geq t \ \Big| \ -\alpha\leq X\leq \alpha  \Big)dt.
\end{equation}
Define the following function: 
\begin{equation}
\mathcal{W}(\alpha,t) \Equaldef \Pr\Big( X \leq \alpha-t \ \Big|\  -\alpha\leq X\leq \alpha  \Big),
\end{equation}
and notice that 
\begin{equation}
\mathcal{W}(\alpha,t) = 0, \ \ t\geq 2\alpha.
\end{equation}
For any fixed $t \in [0,2\alpha]$, the function $\mathcal{W}(\alpha,t)$ is monotone increasing in $\alpha$. In order to show this, consider
\begin{equation}
\mathcal{W}(\alpha,t) = 1- \frac{\Pr\big(\alpha - t \leq X \leq \alpha\big)}{\Pr\big(-\alpha \leq X \leq \alpha\big)}.
\end{equation}
Let
\begin{equation}
f_t(\alpha) \Equaldef \Pr\big(\alpha - t \leq X \leq \alpha\big).
\end{equation}
The fact that $t\in [0,2\alpha]$ implies that the derivative of $f$ with respect to $\alpha$ satisfies:

\begin{equation}
f'_t(\alpha) \leq 0.
\end{equation}

We proceed to define $g$ as:
\begin{equation}
g(\alpha) \Equaldef \Pr\big(-\alpha \leq X \leq \alpha\big).
\end{equation}
It can be easily verified that $g$ is monotone increasing. Therefore, 
\begin{equation}
g'(\alpha)\geq 0.
\end{equation}
Since
\begin{equation}
f_t'(\alpha)g(\alpha) \leq 0 \leq f_t(\alpha)g'(\alpha),
\end{equation}
we have:
\begin{equation}
\left(\frac{f_t(\alpha)}{g(\alpha)}\right)' \leq 0,
\end{equation}
which implies that $\mathcal{W}(\alpha,t)$ is a monotone increasing function of $\alpha$ for all $t \in [0,2\alpha]$. Since 
\begin{equation}
\mathcal{P}(\alpha) = \int_{0}^{2\alpha} \mathcal{W}(\alpha,t)dt
\end{equation}
is a superposition of monotone increasing functions, the function $\mathcal{P}(\alpha)$ is also monotone increasing. The proof of monotonicity of $\mathcal{T}(\alpha)$ follows the same sequence of steps and is omitted for brevity.
\end{IEEEproof}

\section{Proof of Theorem 4}

\begin{IEEEproof}
When the sources are symmetric, we may constrain the optimization of $\mathcal{J}_{q}$ to $a_1 = a_2 = a \in \mathbb{R}$ without loss of optimality. Therefore, 
\begin{equation}
\mathcal{E}^{\mathrm{linear}}_a (1,x) = \begin{bmatrix} x \\ ax 
\end{bmatrix} \ \ \text{and} \ \ \mathcal{E}^{\mathrm{linear}}_{a} (2,x) = \begin{bmatrix}ax \\ x \end{bmatrix}.
\end{equation}
Lemma 2 implies that for all $a<1$, the policy $\mathcal{U}^{\max}$ is optimal for $\mathcal{E}^{\mathrm{linear}}_a$. Evaluating the cost for the pair $(\mathcal{U}^{\max},\mathcal{E}^{\mathrm{linear}}_a)$, we obtain:
\begin{equation}
\mathcal{J}(\mathcal{U}^{\max},\mathcal{E}^{\mathrm{linear}}_a) =  \mathbf{E} \big[(X_2-aX_1)^2\mid U=1\big] \Pr(U=1) + 
 \mathbf{E} \big[(X_1-aX_2)^2\mid U=2\big] \Pr(U=2).
\end{equation}
The symmetry of the sources implies that the first order derivative with respect to $a$ is:
\begin{equation}
\frac{\partial}{\partial a}  \mathcal{J}(\mathcal{U}^{\max},\mathcal{E}^{\mathrm{linear}}_a) = -2 \rho\cdot \sigma^2 +  4a \cdot \mathbf{E}\big[X_1^2 \mid U=1\big]\Pr(U=1).  
\end{equation}
Therefore, the first order optimality condition implies that 
\begin{equation}
a^{\star} = \frac{\rho \cdot \sigma^2}{2\cdot\int_{\mathbb{R}^2}x_1^2 \mathbf{1}(|x_1|\geq |x_2|)f_X(x)dx}.
\end{equation}
It can be shown that $a^{\star}$ computed above satisfies $a^{\star}<1.$
\end{IEEEproof}

\bibliographystyle{IEEEtran}
\bibliography{IEEEabrv,ODSS.bib}

\end{document}